\documentclass[twocolumn,prb]{revtex4-2}
\usepackage{titlesec}
\usepackage{subfigure}
\usepackage{amsmath}
\usepackage{amsfonts}
\usepackage{amssymb}
\usepackage{feynmp}
\usepackage{graphicx}
\usepackage{setspace}
\usepackage{comment}
\usepackage{dsfont}
\usepackage{color}
\usepackage[pdftex,colorlinks=true,linkcolor=blue,citecolor=blue]{hyperref}
\usepackage{float}
\usepackage{enumerate} 
\usepackage{booktabs}
\usepackage[capitalise]{cleveref}
\usepackage{xcolor} 
\newcommand{\ISR}[2]{\mathcal{R}_#2(#1, E)}

\begin{document}
\title{Topological Anderson insulators by latent symmetry}
\author{Jing-Run Lin$^{1}$, Shuo Wang$^{1}$, Hui Li$^{2}$, Zheng-Wei Zuo$^{1}$}
\email{zuozw@haust.edu.cn}
\affiliation{$^1$School of Physics and Engineering, Henan University of Science and Technology, Luoyang 471023, China}
\affiliation{$^2$Department of Finance and Asset Management, Henan University of Science and Technology, Luoyang 471023, China}

\date{\today}

\begin{abstract}
Topological Anderson insulators represent a class of disorder-induced, nontrivial topological states of matter. In this study, we propose a feasible strategy to unveil and design topological Anderson insulators protected by latent symmetries. These are not visible in the original system, but become obvious after performing an isospectral reduction. Using this technique, we design a family of disordered multi-atomic chains that exhibit latent chiral symmetry or mirror (inversion) symmetry. Using topological invariants, bulk polarization, and the divergence of localization length of the topological bound edge states in the reduced disordered system, we show how to identify the gapped and ungapped topological Anderson states in the original systems.  Our work thus extends the concept of topological Anderson insulating phases protected by geometric symmetries and tenfold-way classification to the various types of latent symmetry cases. Overall, our work paves the way for exploiting topological Anderson insulators in terms of latent symmetries.
\end{abstract}

\maketitle

\section{Introduction}
Two main topics in contemporary condensed matter physics are Anderson localization \cite{Anderson58PR,AbrahamsE79PRL,LeePA81PRL,GadeR93NPB,deMouraFABF98PRL,IzrailevFM99PRL} and topological phases of matter \cite{Hasan10RMP, QiXL11RMP, Bernevig13Book, Franz15RMP, AsbothJ16, WenXG17RMP, BergholtzEJ21RMP, MoessnerR21Book}. In recent years, the combination of these two topics has proved to be a fruitful ground for new insights. While topological quantum states are usually robust against weak disorder, sufficiently strong disorder could induce gapped topological Anderson insulators (TAIs) \cite{LiJ09PRL,GrothCW09PRL,GuoHM10PRL, XingYX11PRB, MondragonShem14PRL, AltlandA14PRL2, XiaoZY23PRL}, ungapped TAIs \cite{XuDW12PRB,YangYB21PRB,RenMN24PRL}, Chern TAIs \cite{SuY16PRB}, and topological inverse Anderson insulators \cite{ZuoZW24PRB}.

The physical mechanism of topological edge states in the topological Anderson insulating phases could be explained by effective medium theory \cite{GrothCW09PRL}. In contrast to the band gap of topological insulators, the TAI states are protected by a mobility gap \cite{ZhangYY12PRB}.  In general, TAI with distinct symmetries are characterized by distinct topological invariants.  For example, the winding number classifies the chiral or particle-hole symmetries protected TAIs \cite{MeierEJ18SCI,ZuoZW22PRA,LiuSN22PLA}. Realization of the higher-order TAIs demands the satisfaction of corresponding point group symmetry \cite{YangYB21PRB,ZhangWX21PRL}.  Additionally, the inversion symmetry \cite{VeluryS21PRB}, time-reversal symmetry \cite{LiJ09PRL}, and average symmetry \cite{FulgaIC14PRB, ZhangJ22PRB,MaRC23PRX} could induce and protect the TAI states. Remarkably, the TAIs with different symmetries have been experimentally observed in disordered cold atomic systems \cite{MeierEJ18SCI}, photonic crystals \cite{StuetzerS18NT,LiuGG20PRL,CuiXH22PRL,RenMN24PRL}, and electric circuits \cite{ZhangWX21PRL}.

Despite the significant level of understanding of disorder and topology, the existence of topological Anderson insulator states beyond geometric symmetries and the tenfold-way classification \cite{schnyderClassificationTopologicalInsulators2008,kitaevPeriodicTableTopological2009} remains an open question.  Due to their inherent complexity and the existence of disorder, the conventional topological tools such as symmetry-indicator theory \cite{PoHC17NTC, WatanabeH18SA,KruthoffJ17PRX, Slager12NTP}, topological quantum chemistry \cite{BradlynB17NT,ElcoroL21NTC}, and spin group theory \cite{XiaoZY24PRX, ChenXB24PRX, JiangY24PRX}, cannot characterize and classify the topological features of current systems.  In this study, we propose a new type of topological Anderson insulating phase that is protected by \emph{latent symmetry}. Such a symmetry becomes obvious after performing the isospectral reduction (ISR) technique from graph theory \cite{bunimovichIsospectralTransformationsNew2014,barrettEquitableDecompositionsGraphs2017,godsilStronglyCospectralVertices2017,smithHiddenSymmetriesReal2019,kemptonCharacterizingCospectralVertices2020,morfoniosCospectralityPreservingGraph2021,rontgenDesigningPrettyGood2020,LinJR25CPB,duarteEigenvectorsIsospectralGraph2015a}. On a technical level, the ISR acts as an effective Hamiltonian, effectively diminishing system size while preserving spectral properties, which facilitates the analysis and comprehension of intricate networks and lattice systems.  The ISR has proven applicable to a range of physical systems and problems, including explaining accidental degeneracies \cite{RontgenM21PRL}, designing lattices that host perfectly flat bands \cite{MorfoniosCV21PRB}, finding hidden self-duality in quasiperiodic systems \cite{huHiddenSelfDuality2025}, or topological materials in general \cite{eekFractalityInducedTopology2025,moustajLatentHaldaneModels2025,EekL24PRB,ZhengLY23PRB,Eek24arXiv,RontgenM24PRB,RontgenM23PRL}, to name just a few.

In the following, we utilize the ISR approach to one-dimensional (1D) general tight-binding models with latent symmetries, such that they reduce to the disordered dimerized models.  This framework enables the revelation and characterization of various topological Anderson insulating phases, including both gapped and ungapped TAIs, protected by latent chiral symmetry and mirror symmetry. These phases are identified by topological invariants and the divergence of localization length exhibited by topological edge states. This work elucidates and predicts the existence of latent-symmetry protected topological Anderson insulating phases in 1D disordered multi-atomic chains. 

This paper is organized as follows.  In \cref{MethodModel}, we introduced the ISR method and constructed a general 1D chain through the gluing of building blocks. In \cref{Trimerized}, we take the disordered trimerized model as an example to verify the validity of the ISR approach. The topological Anderson insulators are identified by the topological methods of the reduced disordered Su-Schrieffer-Heeger (SSH) model. In \cref{tetra}, we further investigate the topological Anderson states of the disordered and random flux tetra-atomic chains, which are protected by latent chiral symmetry. In \cref{sec:chiralLatSymm}, how to design a 1D chain with latent chiral symmetry is discussed. In \cref{Inversion}, we analyze the topological properties of the disordered latent mirror-symmetry protected octatomic chain. \cref{Summary} contains our conclusions.

\section{Theoretical foundations}\label{MethodModel}

\subsection{Review of the isospectral reduction}

\begin{figure}[tb]
\centering
\includegraphics[width=0.49\textwidth]{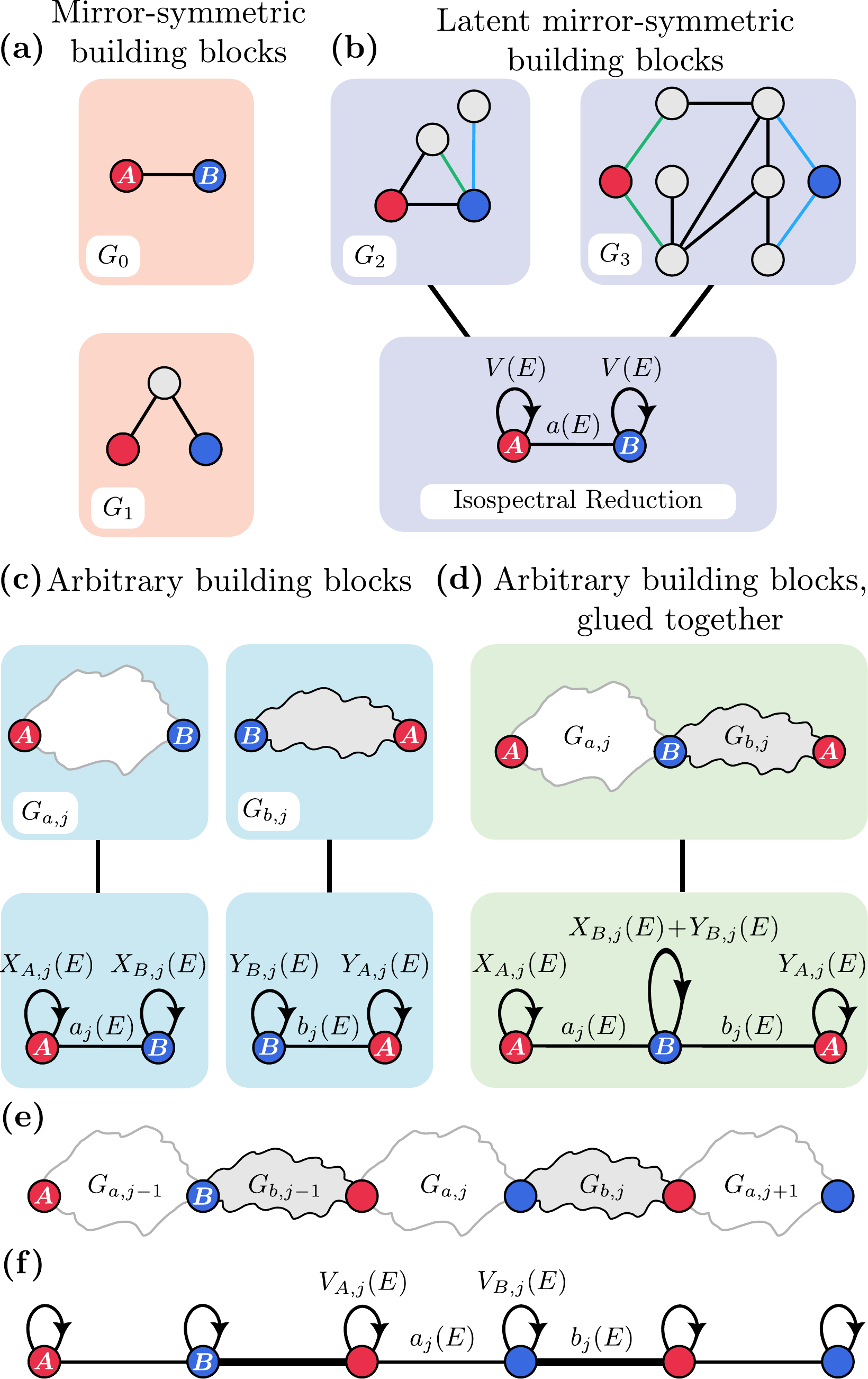}
\caption{
A series of lattice building blocks $G_0$, $G_1$, $G_2$,  and $G_3$ contains lattice sites $S = \{A,B\}$ that are either mirror symmetric (a) or latently mirror symmetric (b).  (c) Two building blocks and their isospectral reduction. The clouds represent an arbitrary system. (d) Gluing together the building blocks from (c) at site $B$ results in a system whose isospectral reduction is directly related to that of the isolated building blocks (see text for details). (e) A one-dimensional chain formed by (possibly different) building blocks, glued together at sites $A$ or $B$. (f) Performing the isospectral reduction over sites $S=\{A,B\}$ yields a one-dimensional chain (not necessarily periodic) of sites $A,B$, with energy-dependent on-site potentials $V_{A,j}$, and $V_{B,j}$, and energy-dependent couplings $a_j(E)$ and $b_j(E)$; see text for details.}
\label{FigBuilding}
\end{figure}

We start by introducing the ISR technique. Starting from an $N$-dimensional Hamiltonian $H$, we partition its $N$ sites into two classes: $S$ (retained lattice sites)  and its complement, $\overline{S}$. With this partitioning, we can write the eigenvalue equation $H \vec{\Psi} = E \vec{\Psi}$ in block-matrix form as
\begin{align}\label{eq0}
\left[
\begin{matrix}
H_{SS} &  H_{S\overline{S}  }   \\
H_{\overline{S}S  } & H_{\overline{S}\overline{S}}  \\
\end{matrix}
\right]
\begin{bmatrix}
\Psi_{S}   \\
\Psi_{\overline{S}  }  \\
\end{bmatrix}
=
E
\begin{bmatrix}
\Psi_{S}   \\
\Psi_{\overline{S}  }  \\
\end{bmatrix} \,,
\end{align}
where $H_{S\overline{S}}$ denotes the submatrix obtained from $H$ by taking the rows in $S$ and columns in $\overline{S}$, and where $\Psi_{S}$ are the components of the full eigenvector $\vec{\Psi}$ on the sites $S$. Writing out \cref{eq0}, we obtain two coupled equations
 \begin{align}   
    H_{SS} \Psi_{S} +H_{S\overline{S}  }\Psi_{\overline{S}  } = E \Psi_{S} ,\label{S}\\
    H_{\overline{S}S  }  \Psi_{S} +H_{\overline{S} \overline{S}  }\Psi_{\overline{S}  } = E \Psi_{\overline{S} } .\label{Sbar}
  \end{align}
From \cref{Sbar}, we obtain
\begin{eqnarray}
 \Psi_{\overline{S} }=(EI - H_{\overline{S}\overline{S}})^{-1}H_{\overline{S}S  }  \Psi_{S}\,,
 \label{PhiSbar}
\end{eqnarray}
with $I$ the identity matrix. 
Inserting Eq.(\ref{PhiSbar}) into \cref{S}, gives us the non-linear eigenvalue problem
\begin{eqnarray} \label{eq:nlevp}
 \left( H_{SS} + H_{S \overline{S} } (EI - H_{\overline{S}\overline{S}})^{-1} H_{\overline{S} S} \right)
\Psi_{S}   
=E\Psi_{S} \,.
\end{eqnarray}
The first term,
\begin{eqnarray}\label{eq00}
\ISR{H}{S} = H_{SS} + H_{S \overline{S} } (EI - H_{\overline{S}\overline{S}})^{-1} H_{\overline{S} S} ,
\end{eqnarray}
is the \emph{isospectral reduction} of the original Hamiltonian $H$ over the (retained) sites $S$ \cite{bunimovichIsospectralTransformationsNew2014}. We note that the ISR is akin to an effective Hamiltonian.

Let us now use the ISR. In the upper panel of \cref{FigBuilding}, we show four different toy models, $G_0$ to $G_3$. The first two structures, $G_0$ and $G_1$, are mirror-symmetric, while there is no obvious geometrical symmetry for the other two structures $G_2$ and $G_3$. However, they feature a latent symmetry. This symmetry becomes obvious after taking the ISR over the red and blue sites $S=\{A,B\}$ (and with $\overline{S}$ the gray sites), which leads to a mirror-symmetric dimer with energy-dependent coupling $a(E)$ and identical on-site potential $V(E)$; see lower panel of \cref{FigBuilding}(b). We note that the exact expression for $a(E)$ and $V(E)$ is different for the two structures.

In matrix form, the ISR of $G_2$---whose $4\times{}4$ Hamiltonian will be given in matrix form in Eq.(\ref{g2}) further below---reads
\begin{equation*}
\mathcal{R}_S(H_{G_2},E)=\left[
\begin{matrix}
J^2/E &  J+J_a J/E   \\
J+J_a J/E & J^2/E  \\
\end{matrix}
\right] \,,
\end{equation*}
and thus $V(E) = J^2/E$ and $a(E) = J+J_a J/E$.

It is important to realize that $G_2$ and $G_3$ are just two examples out of millions of latent symmetric setups that can be easily generated using graph-theoretical tools \cite{rontgenDesigningPrettyGood2020}. Though we invite all readers to read more about the fascinating subject of latent symmetries in the existing literature \cite{smithHiddenSymmetriesReal2019,barrettEquitableDecompositionsGraphs2017,kemptonCharacterizingCospectralVertices2020,morfoniosCospectralityPreservingGraph2021,godsilStronglyCospectralVertices2017}, knowledge of the isospectral reduction is sufficient to comprehend the remainder of this paper.

\subsection{Gluing}
Throughout this paper, we use structures with a (latent) mirror symmetry like $G_0$-$G_3$ as building blocks. By gluing them together, we obtain 1D chains whose isospectral reduction can be conveniently computed using the isospectral reductions of the individual building blocks. This principle is depicted in \cref{FigBuilding}(c) and (d). In \cref{FigBuilding}(c), two different building blocks, $G_{a,j}$ and $G_{b,j}$, along with their isospectral reductions are depicted. Gluing both building blocks together at the site $B$ gives the system depicted in the upper part of \cref{FigBuilding}(d). The isospectral reduction of this composite system is shown in the lower part of \cref{FigBuilding}(d). As can be seen \footnote{
To see this, it is important to keep the notation clear.
To this end, we write the Hamiltonian of $G_{a,j}$ in block-form as
\begin{equation*}
    H_a = \begin{pmatrix}
        H_{xx} & H_{x\overline{S}_a} & H_{xy} \\
        H_{\overline{S}_ax} & H_{\overline{S}_a\overline{S}_a} & H_{\overline{S}_ay} \\
        H_{yx} & H_{y\overline{S}_a} & H_{yy} \\
    \end{pmatrix} \,,
\end{equation*}
where $x:=A$, $y:=B$ are just another label for the two sites over which we reduce when taking the isospectral reduction of $G_{a,j}$, and $\overline{S}_a$ are all the other sites in $G_{a,j}$.
Analogously for $G_{b,j}$, we have
\begin{equation*}
    H_b = \begin{pmatrix}
        H_{uu} & H_{u\overline{S}_b} & H_{uv} \\
        H_{\overline{S}_bu} & H_{\overline{S}_b\overline{S}_b} & H_{\overline{S}_bv} \\
        H_{vu} & H_{v\overline{S}_b} & H_{vv}
    \end{pmatrix} \,,
\end{equation*}
with $u:=A$, $v:=B$, and $\overline{S}_b$ all the other sites in $G_{b,j}$.

When gluing $G_{a,j}$ and $G_{b,j}$ together at the central site $B$ (=$y$ and $u$, respectively), the resulting total Hamiltonian reads (assuming $H_{yy} = H_{uu}$ for simplicity)
\begin{equation*}
    H = \begin{pmatrix}
        H_{xx} & H_{x\overline{S}_a} & H_{xy} & 0 & 0\\
        H_{\overline{S}_ax} & H_{\overline{S}_a\overline{S}_a} & H_{\overline{S}_ay} & 0 & 0 \\
        H_{yx} & H_{y\overline{S}_a} & H_{yy} & H_{u\overline{S}_b} & H_{uv} \\
        0 & 0 & H_{\overline{S}_b u} & H_{\overline{S}_b\overline{S}_b} & H_{\overline{S}_b v} \\
        0 & 0 & H_{vu} & H_{v \overline{S}_b} & H_{vv}
    \end{pmatrix} \,.
\end{equation*}
We then take the isospectral reduction over $S=\{x,y,v\}$, keeping in mind that $y,v$ are identical now.
The relevant matrices are
\begin{equation}
    H_{SS} = 
    \begin{pmatrix}
         H_{xx} & H_{xy} & 0 \\
         H_{yx} & H_{yy} & H_{uv} \\
         0 & H_{vu} & H_{vv}
    \end{pmatrix}\,,
\end{equation}
\begin{equation}
    H_{\overline{SS}} = 
    \begin{pmatrix}
         H_{\overline{S}_a\overline{S}_a} & 0 \\
         0 & H_{\overline{S}_b\overline{S}_b}
    \end{pmatrix}\,,
\end{equation}
\begin{equation}
    H_{S\overline{S}} = 
    \begin{pmatrix}
         H_{x\overline{S}_a} & 0 \\
         H_{y\overline{S}_a} & H_{y\overline{S}_b} \\
         0 & H_{v\overline{S}_b}
    \end{pmatrix}\,,
\end{equation}
and
\begin{equation}
    H_{\overline{S}S} = 
    \begin{pmatrix}
         H_{\overline{S}_a x} & H_{\overline{S}_a y} & 0 \\
         0 & H_{\overline{S}_b y} & H_{\overline{S}_b v}
    \end{pmatrix}\,.    
\end{equation}
Computing the isospectral reduction of only $G_{a,j}$ and $G_{b,j}$ individually, mapping the indices $x,y,u,v$ back to $A$ and $B$, and comparing with the isospectral reduction of the glued Hamiltonian $H$ then yields the result graphically depicted in \cref{FigBuilding}(d).
} after performing the isospectral reduction, the on-site potential of the site $B$ shared by $G_{a,j}$ and $G_{b,j}$ is equal to the sum of the individual on-site potentials of site $B$ in the isospectral reductions of $G_{a,j}$ and $G_{b,j}$.

\subsection{One-dimensional chains}
Using the above gluing principle, we can combine fundamental building blocks such that the isospectral reduction of the resulting composite system is a 1D chain.

In \cref{FigBuilding}(e), we graphically depict this flexible and powerful principle by means of five arbitrary building blocks. Each of them could be, for instance, any of the blocks $G_0$-$G_3$; though it is important to note that the five blocks need not be identical.

Without loss of generality, and for reasons that shall become clear in a moment, we denote the blocks in an alternating manner as ``a'' or type ``b'', and further introduce an index $j$ that enumerates these $a-b$ blocks within the chain.

In \cref{FigBuilding}(f), we depict the isospectral reduction of \cref{FigBuilding}(e). In equation form, the isospectral reduction reads
\begin{eqnarray}\label{eq1}
 \!\!\!\!\!\!\!\mathcal{R}_S(H, E)=[\sum_{j}^{N} a_{j}(E)c^\dagger _{A,j} c_{B,j}
+\sum_{j}^{N -1} b_{j}(E) c^\dagger _{B,j} c_{A,j+1} \notag\\
+h.c.] +\sum_{j}^{N}(V_{A,j}(E) c^\dagger _{A,j} c_{A,j}+ V_{B,j}(E) c^\dagger _{B,j} c_{B,j}) .
\end{eqnarray}
Here, $V_{A,j}(E)$ and $V_{B,j}(E)$ denote the energy-dependent on-site potential of sites $A$ and $B$, respectively, at position $j$ in the chain. The intracell hopping amplitudes are $a_{j}(E)$, and the intercell hopping amplitudes are $b_{j}(E)$. Here, $c^\dagger _{\alpha,j} (c_{\alpha,j})$ denotes the creation (annihilation) operator at site $\alpha$ ($\alpha$ stands for the lattice site $A$ or $B$) at chain position $j$.

Using the above gluing principle, and writing the isospectral reductions of the building blocks as
\begin{equation}\label{eq:Gaj}
    \mathcal{R}_S(G_{a,j},E) = \begin{pmatrix}
        X_{A,j}(E) & a_j(E) \\
        a_j(E) & X_{B,j}(E)
    \end{pmatrix}
\end{equation}
and
\begin{equation}\label{eq:Gbj}
    \mathcal{R}_S(G_{b,j},E) = \begin{pmatrix}
        Y_{A,j}(E) & b_j(E) \\
        b_j(E) & Y_{B,j}(E)
    \end{pmatrix}
\end{equation}
it follows that
\begin{align} \label{eq:onSite1}
    V_{A,j}(E) &= X_{A,j}(E) + Y_{A,j-1}(E) \\
    V_{B,j}(E) &= X_{B,j}(E) + Y_{B,j}(E) \label{eq:onSite2} \,.
\end{align}

\section{Analysis of specific models}

Our main idea for this paper is to build classes of systems whose isospectral reduction is a 1D chain enjoying either chiral [\cref{eq:RSTilde,eq:Gamma,eq:chiralSymmetry}] or mirror [Eqs.(\ref{eq:PSymmetry}) and (\ref{eq:POperator})] symmetry; the original system thus has latent chiral or latent mirror symmetry. As we shall see, this symmetry can be easily achieved using our principle of gluing (latent) mirror symmetric blocks together. We then introduce disorder to the system and explore its topological properties using the isospectral reduction.

In the following, we first give an explicit demonstration of the disordered trimerized model to understand the workflow of unveiling and designing latent-symmetry protected topological Anderson insulators by the ISR approach. Next, we apply the ISR approach to more sophisticated examples.

\subsection{Trimerized chain with chiral symmetry}\label{Trimerized}
\subsubsection{System description}

\begin{figure}[tbp]
\centering
\includegraphics[width=0.48\textwidth]{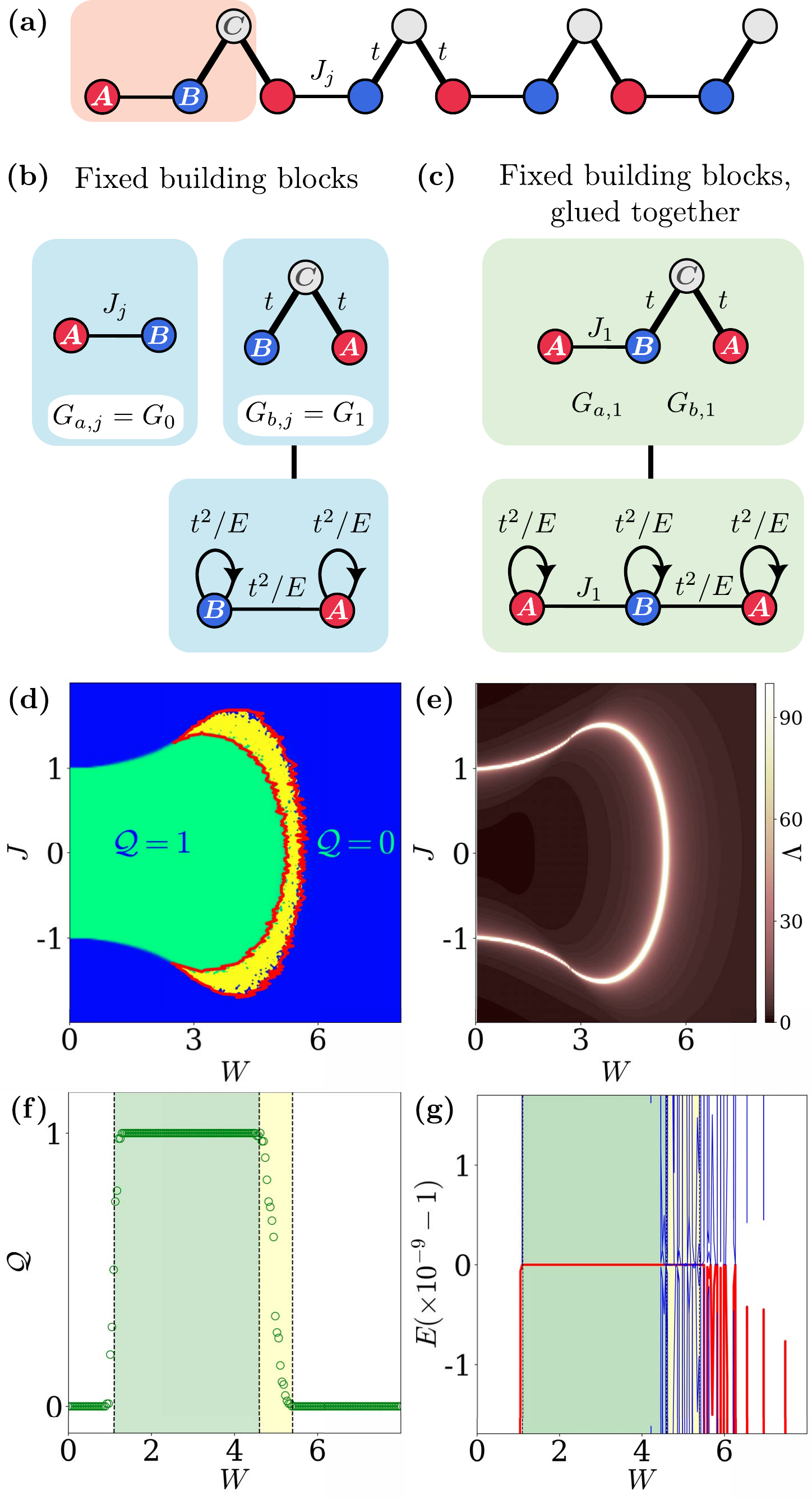}
\caption{
(a) Trimer chain, with sites in the unit cell marked as $A$, $B$, and $C$. The intercell hopping amplitude is $t$, and the intracell hopping amplitudes are $J_j$ and $t$. (b) Two building blocks and isospectral reduction of $G_1$.  (c) Gluing together the building blocks from (b) at site $B$ results in a system. (d) Phase diagram, showing the disorder-averaged real-space topological number $\mathcal{Q}$ in the $W -J$ plane for the disordered trimerized model at filling factor $1/3$ with $t = 1$, and $N=1000$. The green (yellow) region stands for the topological Anderson gapped (ungapped) insulator phase. The blue region indicates the trivial insulator phase.  (e) The divergence of the localization length $\Lambda$ of topological edge states, for the same system parameters as in (d). The white line represents the topological phase boundary. (f) Disorder-averaged topological number $\mathcal{Q}$ and (g) eigenvalues $E$ as a function of $W$ at the initial hopping amplitude $J=1.05$ of (d), where the number of unit cells is fixed as $N=2000$.  Red lines highlight the $N$th eigenvalue, and blue lines show all others. In both (f) and (g), white, green, and yellow shades denote the trivial phase, the gapped TAI phase with bulk energy gap, and the ungapped TAI phase without bulk energy gap, respectively.
}\label{FigTrimer}
\end{figure}

We start by analyzing the rather simple case of a trimerized chain depicted in \cref{FigTrimer}(a), with a unit cell consisting of sites $A,B,C$. To make the connection to our construction principle, we decompose this chain into the two building blocks $G_0$ and $G_1$; cf. \cref{FigBuilding}(a). As one can see, we have a special case of our principle, since all $G_{a,j} = G_0$ are identical (fixed), and also all $G_{b,j} = G_1$. Note that we make the coupling $J_j$ depend on $j$ (that is, on the position in the lattice). This is to let disorder enter the system; we shall do so in a moment.

As shown in Fig.\ref{FigTrimer}(a), the tight-binding model of the trimerized model in real space can be described by
\begin{eqnarray}
H_T=\sum_{j}^{N  } (J_j c^\dagger _{A,j} c_{B,j}+ t c^\dagger _{B,j} c_{C,j})\notag\\
+\sum_{j}^{N -1} t c^\dagger _{C,j} c_{A,j+1}+h.c.\,,\  
\end{eqnarray}
where $c^\dagger _{\alpha,j} (c_{\alpha,j})$ denotes the creation (annihilation) operator at site $\alpha$ ($\alpha$ stands for the lattice sites $A$, $B$, or $C$ in the unit cell) of the $j$th unit cell and $t$ is the intercell hopping amplitude, whereas $t$ and $J_j$ are the intracell hopping amplitudes.

\subsubsection{Isospectral reduction}

Let us start by computing the effective on-site potentials $V_{A,j}(E)$ and $V_{B,j}(E)$ using \cref{eq:onSite1,eq:onSite2}. To this end, we first compute
\begin{equation}
    \mathcal{R}_S(G_{a,j},E) = \begin{pmatrix}
        0 & J_j \\
        J_j & 0
    \end{pmatrix}\,,
\end{equation}
and
\begin{equation}
    \mathcal{R}_S(G_{b,j},E) = \begin{pmatrix}
        t^2/E & t^2/E \\
        t^2/E & t^2/E
    \end{pmatrix} \,.
\end{equation}
Thus, we see that for periodic boundary conditions (PBC)
\begin{equation}
    V_{A,j} = V_{B,j} = V = t^2/E
\end{equation}
are all identical, and the isospectral reduction of the chain reads
\begin{eqnarray}
\mathcal{R}_S(H_T, E)&&=[\sum_{j}^{N} J_j c^\dagger _{A,j} c_{B,j}
+\sum_{j}^{N -1} \frac{t^2 }{E} c^\dagger _{B,j} c_{A,j+1} \notag\\
&&+h.c.] +\sum_{j}^{N} \frac{t^2 }{E}(c^\dagger _{A,j} c_{A,j}+ c^\dagger _{B,j} c_{B,j}) .
\end{eqnarray}

Importantly, all on-site potentials of the ISR become identical. We then continue by analyzing
\begin{equation} \label{eq:RSTilde}
    \widetilde{\mathcal{R}}_S(H_T ,E) :=\mathcal{R}_S(H_T ,E) - V I\,,
\end{equation}
where $I$ is the identity matrix. $\widetilde{\mathcal{R}}_S(H_T ,E)$ is thus a matrix with zero diagonal entries and, as we now show, has a chiral symmetry. To this end, we define the chiral operator
\begin{align} \label{eq:Gamma}
 \Gamma=\sum_{j=1}^{N}{[c_{A,j}^{\dagger }c_{A,j}-c_{B,j}^{\dagger }c_{B,j}]} \,.
\end{align}
It can then be easily shown that $\Gamma$ and the $\widetilde{\mathcal{R}}_S(H_T ,E) $ anti-commute,
\begin{equation} \label{eq:chiralSymmetry}
    \Gamma \,\widetilde{\mathcal{R}}_S(H_T ,E)  = - \widetilde{\mathcal{R}}_S(H_T ,E) \, \Gamma \,.
\end{equation}
In other words, after subtraction of $V I$, the ISR features a chiral symmetry.

\subsubsection{Latent chiral symmetry: General considerations}
As written above, we now introduce disorder to the system by letting
\begin{equation}
    J_j = J+W \omega_j
\end{equation}
where $W$ is the disorder strength, and $\omega_j$ random numbers uniformly distributed in the interval $[-1/2, 1/2]$.

Importantly, after subtracting $V I$, our isospectral reduction is nothing but a disordered SSH chain, though with energy-dependent inter-cell coupling $b_j$. This allows us to conveniently compute the topological properties of our system by simple analogies. Since we will rely on these results throughout this paper, let us first make the discussion in a general manner.

In a conventional disordered SSH chain with chiral symmetry, the topological properties are described by the topological number $\mathcal{Q}$ \cite{FulgaIC11PRB,LonghiS20OL}, which depends on the intra-cell couplings $a_j$ and the inter-cell couplings $b_j$. For a system like ours, whose isospectral reduction becomes an effective, disordered SSH chain, this expression needs to be slightly modified. In particular, both the $a_j$ and the $b_j$ are energy-dependent, and the expression becomes
\begin{eqnarray}\label{Chair-Q}
\mathcal{Q}(E)= \frac{1}{2} \left[1 - \mathrm{sign} \left\{ \prod _{j=1}^{N} a_j^2(E) - \prod _{j=1}^{N} b_j^2(E) \right\} \right].
\end{eqnarray}
In practice, we compute $\mathcal{Q}(E)$ at specific energies $E_f$. Here, $E_f$ denotes the energy of the highest occupied state at filling $f$, defined as the ratio of the number of occupied states to the total number of states.

As we know, the divergence of the localization length for the topological localized zero-energy edge modes is accompanied by the topological phase transition of the disordered SSH model with chiral symmetry \cite{MondragonShem14PRL}. The Schr{\"o}dinger equation for the zero-energy eigenstate $\psi$ in this case reads $({\mathcal{R}}_S(H, E_{f})-V I) \psi=0$. We numerically compute the localization length $\Lambda$ of the eigenvector of the edge states using a numerical transfer matrix method \cite{MondragonShem14PRL,LiuSN22PLA,ZuoZW24PRB}.
For the disordered SSH model, we can obtain the inverse of the localization length, $\Lambda^{-1}$, of the zero-energy mode as
\begin{eqnarray}\label{Lambda-inv}
\Lambda^{-1}=\left|\lim_{N \to \infty} \frac{1}{N} \sum _{j=1}^N  \left(\ln \left|b_{j}(E_{f})\right|-\ln |a_{j}(E_{f})| \right) \right|.
\end{eqnarray}
A detailed derivation of this result is given in Appendix \ref{appendixD}.

\subsubsection{Latent chiral symmetry: Particular case}
For the case of our triatomic chain, we have $b_j:= b = t^2/E$ does not depend on $j$, and $a_j = J_j = J + W \omega_j$. Thus, we obtain
\begin{eqnarray}
\Lambda^{-1}=\left|\lim_{N \to \infty} \frac{1}{N}\sum _{j=1}^N\left( \ln \left|\frac{t^2 }{E_f}\right|-\ln |J +W \omega_j|\right) \right|.
\end{eqnarray}
By applying Birkhoff's ergodic theorem \cite{AltlandA14PRL2,songAIIIBDITopological2014}, the ensemble average can be used to evaluate the inverse of the localization length:
\begin{eqnarray}
\Lambda^{-1}= \left|  \int _{-1/2}^{1/2} \,d\omega (\ln \left|\frac{t^2 }{E_f}\right|-\ln \left| J +W \omega _j\right|) \right|.
\end{eqnarray}
The integrations can be performed explicitly. We analytically obtain {the inverse of the localization length of the topological edge states for the original disordered trimerized chain}
\begin{eqnarray}
\Lambda^{-1}=\left| \ln [\frac{2t^2 \left| 2J -W  \right|^{({ J }/{ W  }-{ 1 }/{ 2  })}}{E_f \left| 2J +W   \right|^{({ J }/{ W  }+{ 1 }/{ 2  })}}] +1\right|.
\end{eqnarray}

We note that the inverse of the localization length (Lyapunov exponent) could also be obtained by Oseledec's multiplicative ergodic theorem \cite{hoffmannComputationalPhysicsSelected1996}. 

\subsubsection{Numerical results}

In the following, we numerically show that the trimerized model with chiral symmetry features a quantized topological number $\mathcal{Q}$ {at filling factor $1/3$}. By simulating the disorder-averaged real-space topological number $\mathcal{Q}$ for the system of $N = 1000$ on the $W-J$ plane over $N_c=100$ disorder realizations, we obtain the topological phase diagram shown in Fig.\ref{FigTrimer}(d). Interestingly, there exists a nontrivial $\mathcal{Q}=1$ phase in the green region with moderate disorder strength, indicating the existence of topological states induced by the disorder. The yellow region selected by the red solid line is the ungapped TAI phase which occupies a certain width. In Fig.\ref{FigTrimer}(d,e), numerical calculations show that the topological phase transition revealed by the abrupt changes of the real-space topological number $\mathcal{Q}$ for PBCs and divergence of the localization length $\Lambda$ for open boundary conditions(OBCs) agree with each other.

In the clean limit ($W=0$), the critical condition for the topological phase transition is $J=\pm 1$. When $J \textgreater 1$, there will be a topological state plateau where $\mathcal{Q}=1$ induced by intermediate disorder. In Fig.\ref{FigTrimer}(f), we show the evolution of $\mathcal{Q}$ over the disorder strength $W$, with $J=1.05$ fixed and where the number of unit cells is fixed as $N = 2000$. This figure clearly reveals that disorder can induce a TAI phase transition at $W\approx 1.1$, where the trivial bulk energy gap---defined by the interval between the $N$th and ($N+1$)th eigenvalues under PBC at 1/3 filling---closes and reopens, thereby signaling the topological phase transition with the topological number $\mathcal{Q}$ changing from zero to one.  The system will remain in the TAI phase until $W\approx 5.4$, and then return to the trivial phase due to strong disorder. 

The disorder-averaged topological number $\mathcal{Q}$ is no longer quantized in the ungapped TAI phase [yellow zones in \cref{FigTrimer}(d),(f), and (g)]. To provide a comprehensive view of this regime, we further analyze the eigenvalues as a function of the disorder strength $W$ in Fig.\ref{FigTrimer}(g), where the number of unit cells is fixed as $N = 2000$. The bulk energy gap exists in the gapped TAI phase, and it is closed in the ungapped TAI phase as the disorder strength increases. In the diagrams of the eigenvalue evolution, the $N$th eigenvalue (closest to $E_n=-1$) is highlighted with red lines, while the remaining eigenvalues are depicted by blue lines. The topological edge states are robust and they are transformed into the localized bulk states under strong disorder. It is worth mentioning that, due to the chiral symmetry of the model [\cref{eq:chiralSymmetry}], the gapped TAI phase obeys the bulk-boundary correspondence \cite{alldridgeBulkBoundaryCorrespondenceDisordered2020,prodanBulkBoundaryInvariants2016}.
For moderate disorder ($4.6 \lesssim  W \lesssim 5.4$), the bulk energy gap vanishes at the yellow shaded region of the ungapped TAI phase, and the ($N-1$)th and ($N+1$)th eigenvalues become degenerate. The topological edge states and the localized bulk states coexist stably in the ungapped TAIs \cite{RenMN24PRL}. For the filling $2/3$, we can use a similar analysis to obtain the topological phase transitions. In short, the numerical simulations demonstrate the existence of the gapped and ungapped TAI phase with chiral symmetry induced by disorder.

We emphasize that one could also utilize the bulk polarization \cite{Resta98PRL,Resta99PRL} to reproduce the phase diagram (Fig.\ref{FigTrimer}(d)). Furthermore, in systems with latent chiral symmetry, the eigenvectors of the isospectral effective Hamiltonian are used to calculate the real-space winding number \cite{songAIIIBDITopological2014,MondragonShem14PRL}. We conclude that the isospectral effective Hamiltonian contains all the topological properties of topological Anderson insulators protected by latent symmetries.

\subsection{Tetra-atomic chains with latent chiral symmetry}\label{tetra}

In the above, we explained the framework of the ISR approach and explored the topological properties of a disordered trimerized chain through this flexible framework. However, the underlying strategy is not limited to such basic examples; it can be applied to more complex systems, as we demonstrate below. 

\begin{figure}[tbp]
\centering 
\includegraphics[width=0.48\textwidth]{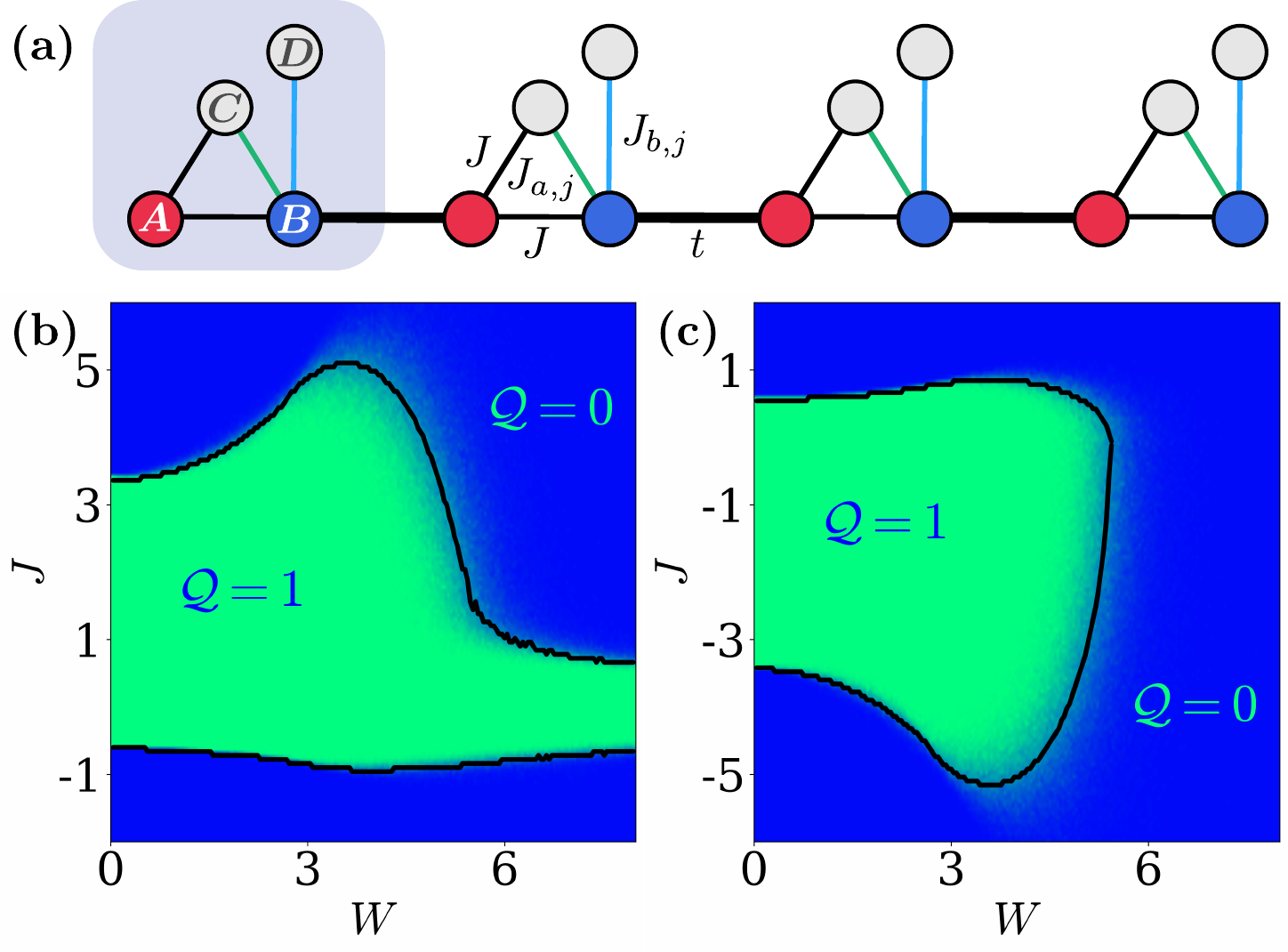}
\caption{(a) Tetra-atomic chain consisting of a four site unit cell, with intracell hopping amplitudes $J$ (black  thin solid line), $J_a$ (green solid line), and $J_b$ (blue solid line). (b,c) Phase diagrams at fillings $1/4$, and $3/4$, respectively, of a tetra-atomic chain with correlated random disorder for intercell hopping $t = 1$.
}\label{FigTetra}
\end{figure}

In particular, let us now investigate TAIs induced by correlated disorder in the tetra-atomic chain illustrated in \cref{FigTetra}(a)]. It consists of $G_2$ as building block for the unit cell, and $G_0$ as the building block for intercell coupling. That is, following our notation of \cref{FigBuilding}(c), we have $G_{a,j}=G_2$, $G_{b,j}=G_0$.
In real space, the tight-binding model thus reads
\begin{eqnarray}\label{HQ1}
H_{Q}=\sum_{j}^{N } [J c^\dagger _{A,j} (c_{B,j}+ c_{C,j})+  c^\dagger _{B,j} (J_{a,j} c_{C,j}\notag\\
+ J_{b,j} c_{D,j})]
+\sum_{j}^{N -1} t c^\dagger _{B,j} c_{A,j+1}+H.c.\,.\  
\end{eqnarray}
where $c^\dagger _{\alpha,j} (c_{\alpha,j})$ denotes the creation (annihilation) operator at site $\alpha$ ($\alpha$ stands for the lattice sites $A$, $B$, $C$, or $D$.) of the $j$th unit cell. The intercell hopping amplitude is $t$, the intracell hopping amplitudes are $J$,$J_{a,j}=J_0+W\omega_j$, and $J_{b,j} =\sqrt{J^2-J_{a,j}^2 }$, where $J_0=J/\sqrt{2}$.
 
The Hamiltonian of the building block $G_2$ is as follows:
\begin{eqnarray}
H_{G_2}=\left[
    \begin{matrix}
        0 & J &J & 0  \\
        J & 0 &J_a& \sqrt{J^2-J_a^2}  \\
        J &J_a  & 0 & 0  \\
        0 &\sqrt{J^2-J_a^2}  & 0 & 0  \\  
    \end{matrix}
    \right],\label{g2}
\end{eqnarray}
Performing the isospectral reductions on the two building blocks yields
\begin{equation}\label{Eq27}
    \mathcal{R}_S(G_{a,j},E) = \begin{pmatrix}
        J^2/E & J+\frac{J (J_0+W\omega_j)}{E} \\
        J+\frac{J (J_0+W\omega_j)}{E} & J^2/E
    \end{pmatrix}\,,
\end{equation}
and
\begin{equation}
    \mathcal{R}_S(G_{b,j},E) = \mathcal{R}_S(G_0,E) = \begin{pmatrix}
        0 & t \\
       t & 0
    \end{pmatrix} \,.
\end{equation}
Thus, using \cref{eq:onSite1,eq:onSite2}, we see that the on-site potentials in the isospectral reduction of our tetra-atomic chain are independent of $j$ and read
\begin{equation}
    V_A = V_B = J^2/E \,.
\end{equation}
Furthermore, we see that the intracell hopping in the isospectral reduction is $a_{j} = J+{J (J_0+W\omega_j)}/{E}$, and the intercell hopping is $b_{j} = t$.

Again, our system thus features a latent chiral symmetry, and we can repeat the same reasoning as done in Sec. \cref{Trimerized}. Thus, the inverse of the localization length for the topological edge states is
\begin{eqnarray}
\Lambda^{-1}=\left|1 +  \ln \left| 2E_f t \frac{ \left| {2x-J W} \right|^{(\frac{x}{J W}-\frac{1}{2})}}{\left| {2x+J W}  \right|^{(\frac{x}{J W}+\frac{1}{2})}} \right| \right|.
\end{eqnarray}
with $x = E_fJ+J J_0$

We investigate the topological numbers $\mathcal{Q}$ of the tetra-atomic chain with fixed $t = 1$ at fillings $1/4$ and $3/4$, which are shown in Figs.\ref{FigTetra}(b,c).  For the Hamiltonian ${\mathcal{R}}_S(H_Q ,E)-V I$, we also plot the divergence points of the localization length $\Lambda$ of zero-energy modes as the black solid lines. The computations of phase diagrams are done for $N=200$ and averaged over $N_c=100$ disorder realizations. These results indicate that the changes of topological number $\mathcal{Q}$ are accompanied by the divergence of localization length, which implies topological phase transitions. In the green region, the system is in the latent chiral symmetry protected topological Anderson insulator phase. 

\begin{figure}[tbp]
\centering 
\includegraphics[width=0.48\textwidth]{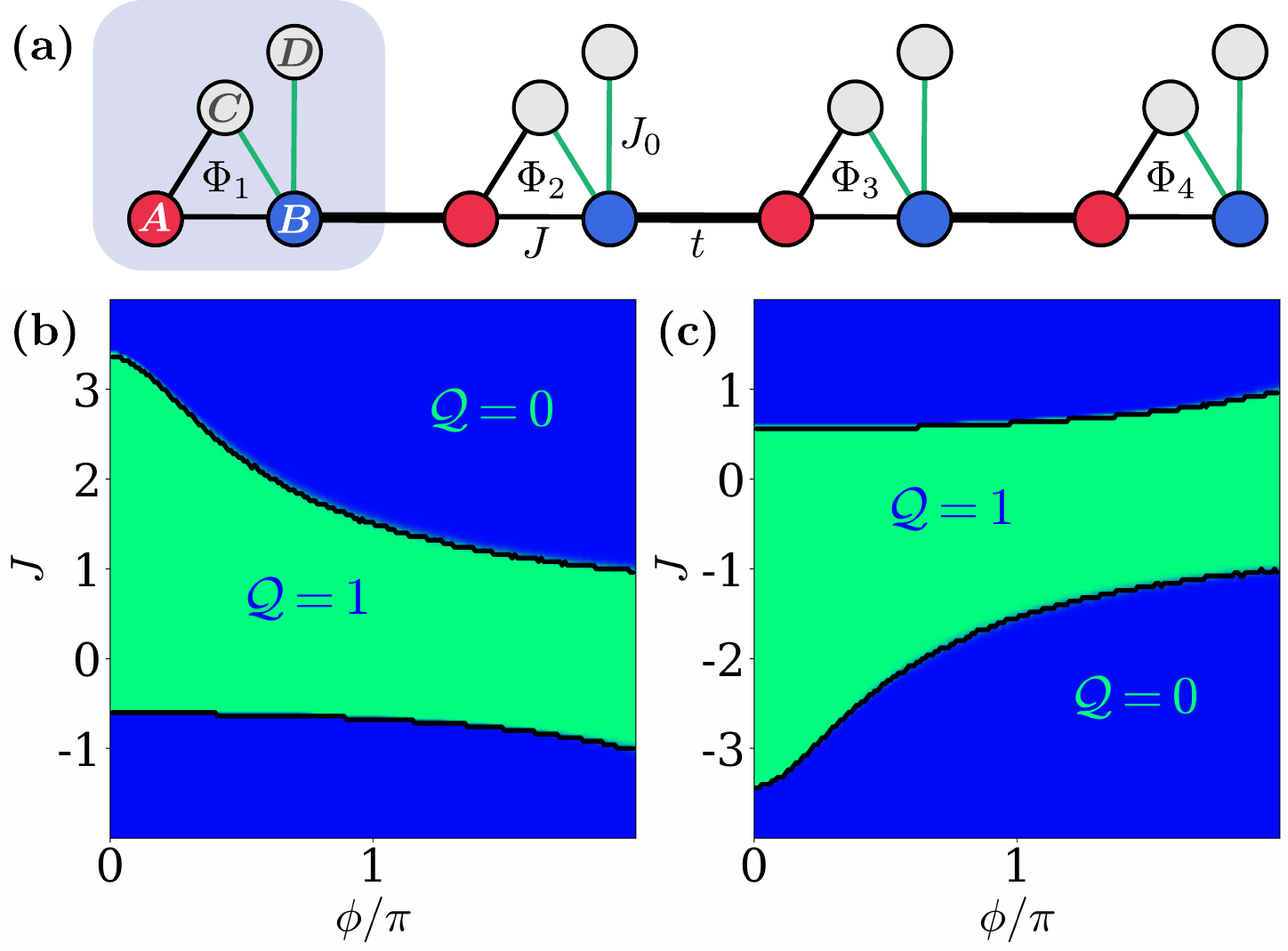}
\caption{
(a) Schematic of a tetra-atomic chain with random flux. (b,c) Phase diagrams at fillings $1/4$, and $3/4$, respectively, of a tetra-atomic chain with random flux and with intercell hopping $t = 1$.
}\label{FigFlux}
\end{figure}

Next, we further investigate the topological Anderson insulator states induced by the random flux in a tetra-atomic chain [illustrated in Fig.\ref{FigFlux}(a)], where the random flux could induce topological phase transitions \cite{LiCA22PRB,liRandomfluxinducedTransitionSequence2025}. This disordered chain could be described by the following tight-binding Hamiltonian
\begin{eqnarray}  
H_{Q}=\sum_{j}^{N } [J e^{i\Phi_j} c^\dagger _{A,j} (c_{B,j}+ c_{C,j})+   J_{0}( c^\dagger _{B,j}c_{C,j}\notag\\
+ c^\dagger _{B,j} c_{D,j})]
+\sum_{j}^{N -1} t c^\dagger _{B,j} c_{A,j+1}+h.c.,\ 
\end{eqnarray}
where $t$ is the intercell hopping amplitude , $J$ is the intracell hopping amplitudes and $J_0=J/\sqrt{2}$. For convenience, we choose the random flux $\Phi_j = \phi \omega_j$ with disorder strength $\phi$ and uniform disorder distribution $\omega_j\in [-1/2, 1/2]$, in units of the magnetic flux quantum $\phi_0=hc/e$.

Performing the ISR on the sites $S=\{A,B\}$, for this disordered tetra-atomic chain model with random flux, the energy-dependent intracell coupling becomes $a_{j} = J e^{i \phi \omega_j} +{J J_0 }/{E}$, the intercell coupling is $b_{j} = t$, and the energy-dependent on-site potential is $V=J^2/E$. Again, this model features a latent chiral symmetry, and we can use Eq. (\ref{Lambda-inv}) to compute the inverse of the localization length for the topological edge states as
\begin{equation*}
\Lambda^{-1}=\left|\lim_{N \to \infty} \frac{1}{N}  \sum _{j=1}^N \left(\ln |t|-\ln \left|J e^{i \phi \omega_j} +\frac{J  J_0 }{E_f} \right|\right)\right| \,.
\end{equation*}

By calculating the disorder-averaged real-space topological number $\mathcal{Q}$ for the system of $N = 200$ on the $W -J$ plane over $N_c=100$ disorder realizations, we obtain the topological phase diagrams shown in Fig.\ref{FigFlux}(b,c). It is easy to see that the random flux can induce TAIs with latent chiral symmetry. In the clean limit of $\Phi = 0$, the system is in the trivial phase with $\mathcal{Q}  = 0$ at filling 1/4 for intracell hopping $J\lesssim -0.6$ ($J\gtrsim  0.6$ at filling 3/4); cf. Fig.\ref{FigFlux}(b) and (c). Interestingly, there exists a nontrivial $\mathcal{Q}  = 1$ phase in the intracell hopping $J\lesssim -0.6$ ($J\gtrsim  0.6$ at filling 3/4) region with moderate disorder strength, indicating the existence of TAIs induced by the random flux.  The black lines in Fig.\ref{FigFlux}(b,c) represent the divergence points of the localization length $\Lambda$ of the zero-energy states of the Hamiltonian matrix $\mathcal{R}_S(H_Q ,E)-V I $, which are in accordance with the topological phase transitions. 

\subsection{Further considerations: Latent chiral symmetry} \label{sec:chiralLatSymm}
In the above, we have investigated several systems with latent chiral symmetry. Let us now briefly discuss how such systems can be designed in a general manner. As a reminder, the isospectral reduction of a latent chiral symmetric system needs to have identical on-site potential $V$ [cf. \cref{eq:RSTilde,eq:Gamma,eq:chiralSymmetry}].

One way to achieve this is to take two arbitrary building blocks $G$ and $G'$ that both feature a (latent) mirror symmetry, and then use $G$ for all the $G_{a,j}$, and $G'$ for all the $G_{b,j}$. The isospectral reduction of a latent mirror symmetric block $G$ has to have the form \footnote{This stems from the fact that the $2\times{}2$ matrix $\mathcal{R}_S(G,E)$ has to commute with $\begin{pmatrix}
    0 & 1 \\ 1 & 0
\end{pmatrix}$ in order to be mirror symmetric.}
\begin{equation}\label{eq:G}
\mathcal{R}_S(G,E)=\left[
\begin{matrix}
V(E) &  a(E)   \\
a(E) & V(E)  \\
\end{matrix}
\right] \,,
\end{equation}
with some $V(E)$ and $a(E)$.
Similarly, the isospectral reduction of $G'$ has to have the form
\begin{equation}\label{eq:G'}
\mathcal{R}_S(G',E)=\left[
\begin{matrix}
V'(E) & b(E)   \\
b(E) & V'(E)  \\
\end{matrix}
\right]
\end{equation}
with some $V'(E)$ and $b(E)$.

Combining these two equations, we see from \cref{eq:onSite1,eq:onSite2} that, for periodic boundary conditions,
\begin{equation} \label{eq:constantDiagonal}
    V_{A,j} = V_{B,j} = V(E) + V'(E)  \,.
\end{equation}
It follows that the system has a latent chiral symmetry.

Another, more involved principle would be to choose the blocks $G$ and $G'$ such that their isospectral reductions have the form
\begin{align*}
\mathcal{R}_S(G,E)&=\left[
\begin{matrix}
V(E) &  a(E)   \\
a(E) & V'(E)  \\
\end{matrix}
\right] \\
\mathcal{R}_S(G',E)&=\left[
\begin{matrix}
V'(E) & b(E)   \\
b(E) & V(E)  \\
\end{matrix}
\right]
\end{align*}
with some $V(E)$, $V'(E)$, $a(E)$, and $b(E)$.
Again, for periodic boundary conditions, \cref{eq:constantDiagonal} would be fulfilled, and thus the system would have a latent chiral symmetry.

 \subsection{Octatomic chain with latent mirror symmetry}\label{Inversion}
 
 \begin{figure}[btp]
\centering 
\includegraphics[width=0.48\textwidth]{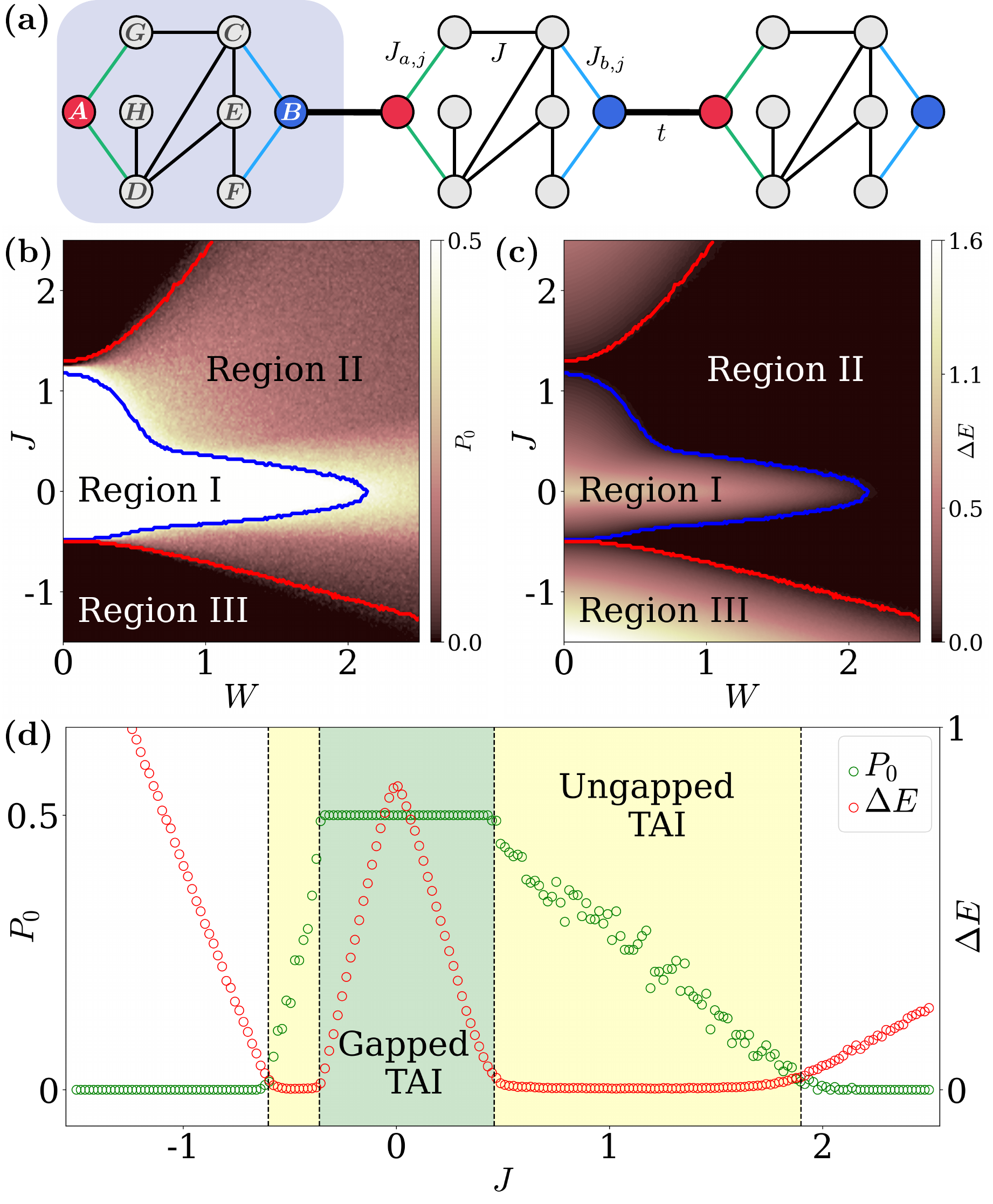}
\caption{(a) Octatomic chain composed of eight lattice points per unit cell. The intercell hopping amplitude is $t$, and the intracell hopping amplitudes are $J$, $J_{a,j}$, and $J_{b,j}$.  (b) Topological invariant $P_0$ and (c) bulk energy gap $\Delta E$ in the $(W, J)$ plane at filling $1/8$ for $t = 1$, $N = 200$, and $N_{c} = 100$. The red (blue) solid line indicates the phase boundary determined by $\Delta E=3 \times 10^{-2}$. (d) Energy gap $\Delta E$ and bulk polarization $P_0$ as a function of $J$ for the case when $N = 200$, $N_c=100$, $t=1$, and $W=0.65$.}\label{FigOctatomic}
\end{figure} 
Having established the features of the multi-atomic chain with latent chiral symmetry and its topological invariants in position space, we now discuss the effects of disorder on a multi-atomic chain that features latent mirror symmetry. As shown in Fig.\ref{FigOctatomic}(a), the lattice building block for intracell is $G_3$, and $G_0$ is the lattice building block for intercell, namely $G_{a,j}=G_3$, $G_{b,j}=G_0$ in Figs.\ref{FigBuilding}(a, b). This Hamiltonian of octatomic chain in real space could be written as
\begin{align}
H_{O}=&\sum_{j}^{N } [J_{a,j} c^\dagger _{A,j} (c_{D,j}+ c_{G,j})+  J_{b,j} c^\dagger _{B,j} (c_{C,j} + c_{F,j})\notag\\
&+J c^\dagger _{C,j} (c_{D,j}+ c_{E,j}+ c_{G,j})+J c^\dagger _{D,j} (c_{E,j}+ c_{H,j})\notag\\
& +J c^\dagger _{E,j} c_{F,j}]+\sum_{j}^{N -1} t c^\dagger _{B,j} c_{A,j+1}+h.c.\,,  
\end{align}
where $c^\dagger _{\alpha,j} (c_{\alpha,j})$ denotes the creation (annihilation) operator at site $\alpha$ ($\alpha$ stands for the lattice sites $A$, $B$, $C$, $D$, $E$, $F$, $G$, or $H$.) of the $j$-th unit cell. The intercell hopping amplitude is $t$, the intracell hopping amplitudes are $J$,$J_{a,j}=J+W\omega_{a,j}$, and $J_{b,j}=J+W\omega_{b,j}$. 

Repeating the same steps as in the previous examples and performing the isospectral reduction over $S=\{A,B\}$, we find that the intracell hopping $a_{j}$, and onsite potential $V_{A,j}, V_{B,j}$ of the resulting effective disordered dimerized chain are given by
\begin{eqnarray}
\left\{
\begin{aligned}
&a_{j} =\frac{2 E^2 J J_{a,j} J_{b,j}}{E^4-E^3 J-4 E^2 J^2+E J^3+J^4} \\
&b_{j} = t \\
&V_{A,j}=\frac{2 E J_{a,j}^2 (E^2-E J-J^2)}{E^4-E^3 J-4 E^2 J^2+E J^3+J^4}\\
&V_{B,j}=\frac{2 E J_{b,j}^2 (E^2-E J-J^2)}{E^4-E^3 J-4 E^2 J^2+E J^3+J^4}\\
\end{aligned}
\right. .
\label{conv}
\end{eqnarray}

For $W=0$, the isospectral reduction of the system is mirror symmetric \footnote{And, additionally, also chiral symmetric.}. That is, under PBC, the isospectral reduction commutes with the mirror operator $\mathcal{P}$,
\begin{equation} \label{eq:PSymmetry}
    \mathcal{P} \,\mathcal{R}_S(H ,E) = \mathcal{R}_S(H ,E) \, \mathcal{P}\,,
\end{equation}
with
\begin{equation}\label{eq:POperator}
    \mathcal{P}=\sum_{j=1}^{N}{[c_{A,N+1-j}^{\dagger}c_{B,j}+c_{B,N+1-j}^{\dagger } c_{A,j}]} \,.
\end{equation} 
Before continuing, and in the same spirit as \cref{sec:chiralLatSymm}, we remark that it is easy to design other systems whose isospectral reduction is a 1D mirror-symmetric chain [that is, which fulfills \cref{eq:PSymmetry}]. Again, this can be achieved by taking two arbitrary building blocks $G$ and $G'$ that both feature a (latent) mirror symmetry, and then using $G$ for all the $G_{a,j}$, and $G'$ for all the $G_{b,j}$. As discussed in \cref{sec:chiralLatSymm}, under PBC, all on-site potentials of the isospectral reduction of this system are identical. Furthermore, all $b_j$ are identical and independent of $j$, and the same applies to the $a_j$. As a result, the isospectral reduction of the system over $S=\{A,B\}$ will be mirror-symmetric.

Going back to our system at hand, we now introduce disorder that preserves the mirror symmetry. That is, we restrict the random numbers by $\omega_{a,j}=\omega_{b,N+1-j}$ for each disorder realization, such that the onsite potentials $V_{A,j}=V_{B,N+1-j}$ and intracell hopping amplitudes $a_{j}=a_{N+1-j}$ are constrained to ensure the mirror symmetry. The reduced effective Hamiltonian ${\mathcal{R}}_S(H_{O},E)$ then takes on the mathematical form of a disordered mirror-symmetric chain. 

Since the octatomic chain is latently mirror-symmetric, we can characterize the topological phase transitions of the setup using the polarization \cite{Resta98PRL} in real space as the topological invariant. It is defined as
\begin{eqnarray}
 P_{0} = \sum_{n}^{N} \left( \frac{1}{2\pi} {\rm Im }\ln \xi _n \right) \,. 
 \end{eqnarray}
where the set of $\{\xi _n\}$ is the eigenvalues of the matrix $X_P = P_{occ}X P_{occ}$, where $X=\sum_{j}^{N}  e^{\frac{2\pi i}{N}j}(|{A,j} \rangle  \langle {A,j} | + | {B,j} \rangle \langle {B,j} |)$ is the position operator, $P_{occ} = \sum_{occ.} |\psi_i \rangle  \langle \psi_i | $  is the projector onto the subspace of occupied states, with $|\psi_i\rangle$ being the eigenstate of $\mathcal{R}_S(H, E)$ with energy $E_i$. We remind the reader that, according to Eqs.(\ref{eq:nlevp}) and (\ref{eq00}), $\psi_{i}$ is given by the $S$-components of the corresponding eigenvector of the original system (at the same energy $E$).

In Fig.\ref{FigOctatomic}(b), we categorize the topological phases region into two distinct phases at $1/8$ filling: the gapped TAIs in region I and the ungapped TAIs in region II. This categorization is based on whether the $N$th and $(N + 1)$th eigenvalues are degenerate. For each disorder realization denoted by the label $s$, the corresponding disorder-averaged bulk gap $\Delta E$ under PBC is given by
\begin{eqnarray}
\Delta E=\frac{1}{N_{c}}\sum_{n=1}^{N_{c}} |E_{s,N + 1}-E_{s,N}| .
\end{eqnarray} 
From this we plot the disorder-averaged energy gap on the $W-J$ plane, shown in \cref{FigOctatomic}(c).

Figure \ref{FigOctatomic}(b) illustrates that the disorder-averaged value of the polarization $P_0$ is quantized in region I and non-trivial up to the boundary of fluctuation onset value (blue solid line). This is because the bulk polarization fluctuates between a set of quantized values which alters the disorder-averaged value of bulk polarization in the ungapped TAI region. Comparing Fig.\ref{FigOctatomic}(b) and Fig.\ref{FigOctatomic}(c), it is obvious that the bulk polarization $P_0$ fluctuations are connected to the closing of the disorder-averaged energy spectral gap. When the bulk energy spectral gap closes, the disorder-averaged value of the bulk polarization $P_0$ is no longer quantized in region II of the phase diagram. In other words, as long as the energy gap does not close, the system still remains in the gapped TAI state. The disorder-averaged polarization invariant deviates from its quantized value in the ungapped region due to the exchange of occupied and unoccupied states at mirror-symmetric centers at $1/8$ filling \cite{VeluryS21PRB}. It should be noted that the bulk polarization $P_0$ for a single disorder realization is still quantized and well-defined for localized bulk states in disordered systems \cite{Resta99PRL}. When the system evolves across the phase boundary (red solid line), there is a change of polarization $P_0$ value becoming zero. Beyond the blue solid line, the region within the red solid line corresponds to the energy gap $\Delta E\leqslant 3 \times 10^{-2}$. In region III of Fig.\ref{FigOctatomic}(b), the system is in a trivial phase and the energy gap opens again.

Hereafter, we take the evolution of polarization $P_0$ and bulk gap $\Delta E$ with couplings $J$ at $W=0.65$ as an example. As shown in Fig.\ref{FigOctatomic}(d), we categorize the region of $W=0.65$ into three distinct phases:  the trivial insulators ($J \lesssim-0.6 $ and $ 1.9 \lesssim  J$, white region), the gapped TAIs ($-0.36 \lesssim  J \lesssim  0.46$, green shaded region) and the ungapped TAIs ($-0.6 \lesssim  J \lesssim -0.36$ and $0.46 \lesssim  J \lesssim  1.9$, yellow shaded region). When $J \approx  -0.36$, the polarization changed progressively from $P_0=0$ to $P_0=1/2$, which clearly reveals that disorder induces the topological phase transition. The system will remain in the TAI phase until $J \approx  1.9$, and then return to the trivial phase. The computations of Fig.\ref{FigOctatomic}(d) were done for $N=200$ and averaged over $N_c=100$ disorder realizations. From the above discussion, we show the possibility of the disorder-driven topological phases protected by latent mirror symmetry including the gapped and ungapped TAIs.

\section{Summary and outlook}\label{Summary}

In short, we propose a theoretical framework leveraging the isospectral reduction approach from graph theory to uncover topological Anderson insulator states protected by latent symmetries in disordered multi-atomic chains. Taking the tetra-atomic chain and octatomic chain as representative examples, we demonstrate that this reduction approach simplifies complex lattice systems of many degrees of freedom while preserving their essential topological features. Latent symmetries and topological properties emerge clearly within the reduced systems, enabling characterization of the original systems' topological phases through topological invariants and edge states. Furthermore, these latent symmetry protected topological phases can be directly verified in current experimental techniques such as cold atoms \cite{MeierEJ18SCI}, photonic crystals \cite{StuetzerS18NT}, and topolectrical circuits \cite{ZhangWX21PRL}. Our work provides valuable insights for discovering and designing the latent symmetry protected topological Anderson insulators.

\emph{Acknowledgments.}---We thank Malte R{\"o}ntgen for useful discussions. This work was partially supported by the National Natural Science Foundation of China (Grants No. 12574010 and No. 12074101) and the Natural Science Foundation of Henan (Grant No. 212300410040).

\appendix
\begin{widetext}
 \section{Localization length}\label{appendixD}
From the tight-binding Hamiltonian of 1D disordered SSH chain, we directly read off the transfer matrix of the zero-energy mode ${{\mathcal{M}}_{j}}=\left( \begin{matrix}
   -\frac{{{a}_{j}}}{{{b}_{j}}} & 0  \\
   0 & -\frac{{{b}_{j}}}{{{a}_{j+1}}}  \\
\end{matrix} \right)$ and $\left( \begin{matrix}
   {{\psi }_{A,j+1}}  \\
   {{\psi }_{B,j+1}}  \\
\end{matrix} \right)=\text{ }{{\mathcal{M}}_{j}}\left( \begin{matrix}
   {{\psi }_{A,j}}  \\
   {{\psi }_{B,j}}  \\
\end{matrix} \right)$. 
Multiplying the transfer matrices, we could find
\begin{eqnarray} \left( \begin{matrix}
   {{\psi }_{A,N}}  \\
   {{\psi }_{B,N}}  \\
\end{matrix} \right)=\text{ }{{\mathcal{M}}_{N}}{{\mathcal{M}}_{N-1}}\cdots {{\mathcal{M}}_{2}}{{\mathcal{M}}_{1}}\left( \begin{matrix}
   {{\psi }_{A,1}}  \\
   {{\psi }_{B,1}}  \\
\end{matrix} \right)=\left( \begin{matrix}
   {{(-1)}^{N}}\prod\limits_{j}^{N}{\frac{{{a}_{j}}}{{{b}_{j}}}} & 0  \\
   0 & {{(-1)}^{N}}\prod\limits_{j}^{N}{\frac{{{b}_{j}}}{{{a}_{j+1}}}}  \\
\end{matrix} \right)\left( \begin{matrix}
   {{\psi }_{A,1}}  \\
   {{\psi }_{B,1}}  \\
\end{matrix} \right). 
\end{eqnarray} 
Taking ${{\psi }_{1,A}}={{\psi }_{1,B}}=1$, we can get the inverse of the localization length 
\begin{eqnarray} {{\Lambda }^{-1}}=\max \{\underset{N\to \infty }{\mathop{\lim }}\,\frac{1}{N}\ln |{{\psi }_{N,A}}|,\underset{N\to \infty }{\mathop{\lim }}\,\frac{1}{N}\ln |{{\psi }_{N,B}}|\}.
\end{eqnarray} 
 Then, we can obtain (in the thermodynamic limit $N\to \infty $)
\begin{eqnarray} {{\Lambda }^{-1}}=|\underset{N\to \infty }{\mathop{\lim }}\,\frac{1}{N}\sum\limits_{j=1}^{N}{(\ln |{{b}_{j}}|-\ln |{{a}_{j}}|)}|.
\end{eqnarray} 
\end{widetext}

\end{document}